\documentstyle[12pt,aaspp4]{article}
\lefthead{Oka et al.}
\righthead{A DARK CLOUD ASSOCIATED WITH AN UNIDENTIFIED EGRET SOURCE}

\def\kms{\hbox{km s$^{-1}$}}
\def\VLSR{\hbox{$V_{\rm LSR}$}}
\def\hh{\hbox{$^{\rm h}$}}
\def\mm{\hbox{$^{\rm m}$}}
\def\ss{\hbox{$^{\rm s}$}}

\begin{document}

\title{A DARK CLOUD ASSOCIATED WITH AN UNIDENTIFIED EGRET SOURCE} 

\author{Tomoharu Oka\altaffilmark{1,2}, Nobuyuki Kawai\altaffilmark{2}, Tsuguya Naito\altaffilmark{3}, 
Takahiko Horiuchi\altaffilmark{4,2}, Masaaki Namiki\altaffilmark{5,2}, Yoshitaka Saito\altaffilmark{6,2}, 
Roger W. Romani\altaffilmark{7}, and Tadashi Kifune\altaffilmark{8} }

\altaffiltext{1}{{\footnotesize Department of Physics, Faculty of Science, The University of Tokyo, 7-3-1 Hongo, Bunkyo-ku, Tokyo 113-0033, Japan.}} 
\altaffiltext{2}{{\footnotesize Cosmic Radiation Laboratory, The Institute of Physical and Chemical Research (RIKEN), 2-1 Hirosawa, Wako, Saitama 351-0198, Japan.}} 
\altaffiltext{3}{{\footnotesize National Astronomical Observatory, 2-21-1 Osawa, Mitaka, Tokyo 181-8588, Japan.}} 
\altaffiltext{4}{{\footnotesize Department of Physics, Faculty of Science, Chiba University, 1-33 Yayoicho, Inage, Chiba, Chiba 263-0022, Japan.}} 
\altaffiltext{5}{{\footnotesize Department of Physics, Faculty of Science, Science University of Tokyo, 1-3 Kagurazaka, Shinjyuku, Tokyo 162-8601, Japan.}} 
\altaffiltext{6}{{\footnotesize Institute of Space and Astronautical Science (ISAS), 3-1-1 Yoshinodai, Sagamihara, Kanagawa 229-0022, Japan.}} 
\altaffiltext{7}{{\footnotesize Department of Physics, Stanford University, Stanford, CA 94305-4060, U.S.A.}} 
\altaffiltext{8}{{\footnotesize Institute for Cosmic Ray Research (ICRR), University of Tokyo, 3-2-1, Midoricho, Tanashi, Tokyo 188-8502, Japan.}}

\authoremail{tomo@taurus.phys.s.u-tokyo.ac.jp}

\begin{center}
{Accepted for publication in ApJ}
\end{center}

\begin{abstract}
CO {\it J}=1--0 mapping observations of the dark nebula Lynds 227, 
which is possibly associated with the unidentified EGRET source 2EG J1811--2339, are presented. 
We detected a large amount of molecular gas along Lynds 227 with a total mass of (1--2)$\times 10^4\; M_{\sun}$ 
surrounding an X-ray synchrotron nebula, 
which was detected by the ASCA satellite within the error circle of 2EG J1811--2339. 
Molecular gas along Lynds 227 shows spatial anti-correlation with the X-ray synchrotron nebula, 
suggesting that the synchrotron nebula is interacting with the Lynds 227 cloud. 
We propose a $\gamma$-ray emission mechanism for 2EG J1811--2339: 
High energy electrons are injected from a rotation-powered pulsar 
and accelerated further in shock waves generated by the interaction with ambient matter. 
The high energy electrons move into the molecular cloud at the Lynds 227 
and collide with dense interstellar matter to yield high energy $\gamma$-ray photons, 
mainly through relativistic bremsstrahlung. 
\end{abstract}

\section{INTRODUCTION}

The EGRET instrument on the Compton Gamma Ray Observatory detected several tens of 
high energy ($E\!>\!100$ MeV) $\gamma$-ray sources (Thompson et al. 1995, 1996), 
many of which are unidentified at other wavelengths. 
35 EGRET sources detected with high confidence concentrate along the Galactic plane ($|b|\!<\!10$\arcdeg ). 
Six of them have now been identified from their pulsations as young neutron stars. 
Although the other sources are most likely rotation-powered pulsars as well, 
some sources in this category show time variability which is never seen in the $\gamma$-ray emitting pulsars. 
%and their number and spatial distribution may not be consistent with pulsars. 
Statistical studies showed that the low-latitude unidentified EGRET sources could be associated with 
pulsars, supernova remnants, or OB associations 
(Yadigaroglu \&\ Romani 1995; Sturner \&\ Dermer 1995; Yadigaroglu \&\ Romani 1997), 
neither direct associations nor identifications at other wavelengths have been confirmed, however.

2EG J1811--2339 is one of the low-latitude unidentified EGRET sources detected with high significance above 1 GeV; 
this leads to a relatively small 95\%\ error circle of radius 13$^\prime$ (Lamb and Macomb 1997). 
Unlike the $\gamma$-ray pulsars, such as Vela, PSR1055-52 and PSR1706-44, 
$\gamma$-ray emission from 2EG J1811--2339 seems to be time variable (McLaughlin et al. 1996), 
which argues against a magnetospheric origin. 
Kawai et al. (1998) have performed X-ray observations toward 2EG J1811--2339 with ASCA satellite 
to search for a rotation-powered pulsar within the error circle. 
They detected an X-ray nebula with nonthermal spectrum and a group of point like sources 
within the error circle of 2EG J1811--2339. 
These point like sources exhibit spectra with moderate absorption columns 
[$n({\rm H})\!=\!\mbox{(4--6)}\times 10^{21}$ cm$^{-3}$] and X-ray luminosities of $\sim\!10^{33}$ erg s$^{-1}$. 
This group of X-ray point like sources resembles a star forming region (Koyama et al. 1997). 
Actually, the location of an X-ray source is coincident with a stellar cluster in an HII region (Sharpless 32), 
which is associated with a dark nebula (Lynds 227).

However, the hard X-ray synchrotron nebula implies the existence of an embedded pulsar,
because the extremely high-energy electrons (with $E\!\sim$ several TeV) responsible for the synchrotron emission 
(in magnetic field of $B \sim {\rm mG}$) are most likely to be supplied by a rotation-powered pulsar. 
This supports a pulsar-related mechanism for the $\gamma$-ray emission from 2EG J1811--2339. 
These results are summarized as follows: 
(1) The X-ray synchrotron nebula suggests the existence of high energy charged particles, 
which may be injected from a rotation-powered pulsar. 
(2) The absorption columns toward X-ray point sources suggest that these sources are behind 
or embedded in a region with rather high density such as an interstellar molecular cloud. 
Other evidence for this cloud comes from an HII region (S32) and a dark nebula (L227) 
within the error circle of 2EG J1811--2339. 
(3) Time variation of 2EG J1811--2339 suggests that some of the 
$\gamma$-ray emission are not directly from the pulsar magnetosphere.

From these results, we propose the following 
high energy $\gamma$-ray emission mechanism for 2EG J1811--2339.  
A relativistic wind of charged particles, injected as a pulsar wind, 
interacts with dense interstellar matter in a molecular cloud and shock waves are generated.  
In the pulsar wind termination shock the particles are accelerated by Fermi mechanism 
to convert into a power law spectrum 
(see reviews;  Drury 1983; Blanford \& Eichler 1987; Jones \& Ellison 1991).
The accelerated high energy electrons provide X-ray synchrotron nebula,
which resembles Crab (Kennel \& Coroniti 1984a; 1984b).
The high energy particles may move into molecular cloud 
to produce high energy $\gamma$-ray photons 
through relativistic bremsstrahlung process.  
Instabilities in the region, 
where the pulsar wind interacts with the molecular cloud, 
and inhomogeneous target gas may cause variable $\gamma$-ray emission on $\sim yr$ time scale.

Physical parameters of the interstellar matter around the pulsar 
as well as the evidence for the direct association may be crucial 
for verifying the above hypothesis for $\gamma$-ray emission. 
We performed CO {\it J}=1--0 observations toward the Lynds 227 dark nebula 
using the Nobeyama Radio Observatory 45 m radio telescope. 
The {\it J}=1--0 rotational transition of the common isotopic species of carbon monoxide, 
$^{12}$C$^{16}$O is the most often used tracer of dense 
($n({\rm H}_2)\!\gtrsim\!10^2$ cm$^{-3}$) interstellar matter in molecular phase.  
We present in this paper the results of our CO {\it J}=1--0 observations (\S 3), 
examining the possibility of the direct association of the pulsar with the L227 cloud (\S 4.1), 
and discuss the $\gamma$-ray emissivity from the L227/pulsar system (\S 4.2).

\section{OBSERVATIONS}

CO {\it J}=1--0 (115.271204 GHz) observations of L227 have been made 
with the NRO 45 m telescope in February 1998. 
The NRO 45 m telescope has a $17\arcsec\!\pm\! 1\arcsec$ full width at half maximum beam at 110 GHz. 
%The main beam efficiency of the telescope is $0.52\pm 0.06$ at 110 GHz (S100). 
We employed the $2\!\times\!2$ focal-plane array SIS receiver S115Q. 
The system noise temperature was 400--600 K during the observations. 
Calibration of the antenna temperature was accomplished by chopping 
between an ambient temperature load and the sky.  
Since the mixers in S115Q work in a quasi-single sideband mode, 
we scaled the antenna temperature for each channel referring to 
that taken with the single-beam SIS receiver S100 
equipped with a quasi-optical image rejection filter. 
Every day, we observed the molecular core in the Lagoon nebula (White et al. 1997), 
at $(\alpha_{1950}, \delta_{1950}) = (+18\hh 00\mm 36.3\ss , -24\arcdeg 22\arcmin 53\arcsec )$ 
near its transit with each channel of S115Q, and calculated scaling factors 
by comparing with $T_A^{*}({\rm S100})\!=\!23.7$ K. 
We used the high-resolution acousto-optical spectrometers (AOS-Hs) 
each of which cover an instantaneous bandwidth of 40 MHz with a spectral resolution of 37 kHz. 
At the $^{12}$CO frequency, these correspond to 104 km s$^{-1}$ velocity coverage and 0.01 \kms\ 
velocity resolution, respectively.

All data were obtained by position switching between target positions and the reference position, 
$(\alpha_{2000}, \delta_{2000})\!=\!(+18\hh 09\mm 20\ss , -23\arcdeg 37\arcmin 00\arcsec )$. 
The reference position was carefully checked and no CO emission exceeding 0.3 K was detected. 
To conduct the survey effectively, three on-source positions were observed for one reference position observation. 
The spectra were obtained with two grid spacings. 
The coarse grid survey has a spacing of 34\arcsec, and covers almost the full extent of the cloud; 
the fine grid survey has a spacing of 17\arcsec, and covers the area around the X-ray source. 
In the fine grid survey, the centers of the diagonals of 17\arcsec\ grid are also filled 
for CO emitting areas to pick up the emission fully.
%We also took $^{13}$CO {\it J}=1--0 spectra for the area adjacent to the X-ray synchrotron nebula 
%in a 17\arcsec\ spacing.
Typically 40 seconds on source integration for the fine grid survey gave spectra with $\sigma\!\leq\! 0.5$ K 
at a velocity resolution of 0.01 km s$^{-1}$, while 20 seconds integration 
for the coarse grid survey gave spectra with $\sigma\!\leq\! 1.0$ K. 
Pointing errors were corrected every two hours by observing the SiO maser source VX Sgr 
with the HEMT amplifier receiver H40. 
The pointing accuracy of the telescope was good to $\leq\! 3\arcsec$ in both azimuth and elevation.

The data were reduced on the NEWSTAR reduction package. 
We subtracted baselines of the spectra by fitting third degree polynomial lines. All the CO data 
were calibrated in units of corrected antenna temperature $T_R^{*}$ ($\!=\!T_A^{*}/\eta_{\rm fss}$). 
The forward spillover and scattering efficiency $\eta_{\rm fss}$ of the 45 m telescope is 0.58 (Oka et al. 1998).

\section{RESULTS}
\subsection{Molecular Gas Distribution}

In Fig.1, a map of the integrated emission in the CO {\it J}=1--0 line is shown 
for the area covered by the coarse grid survey. 
Although faint emission extends outside of the mapping area, 
the coarse grid map covers the main ridge of the molecular cloud which corresponds to L227. 
Molecular gas occupies an arc with a radius of $\sim\!4$ pc encircling S32 and the X-ray point sources, 
showing a good correlation with the dark nebula L227 seen in obscuration of the background star light. 
The cloud has a remarkably sharp emission rim on the side facing S32 and the X-ray synchrotron nebula, 
compared to rims on the other side.
The map of CO {\it J}=1--0 integrated emission of the area covered by the fine grid survey is shown in Fig.2. 
%The mean $^{12}$CO/$^{13}$CO intensity ratio over the $^{13}$CO mapping area is 4.9, 
%giving the mean $^{12}$CO opacity as high as $13$ %if we adopt [$^{12}$CO]/[$^{13}$CO] $\!=\!57$ (Langer \&\ Penzias 1990). 

Fig.3 shows the CO {\it J}=1--0 velocity channel maps for the area covered by the fine grid survey. 
The main component of the L227 cloud has velocities $\VLSR\!=\!7$--$11$ \kms, 
which is similar to the H$\alpha$ velocity of S32, $\VLSR\!=\!6.2\!\pm\! 2.0$ \kms\ (Fich, Treffers, \&\ Dahl 1990). 
We also see another small isolated clump at $\VLSR\!\simeq\! 19$ \kms, velocity-separated from the main component. 
This clump is spatially adjacent to the X-ray synchrotron nebula in the plane of the sky, 
direct association of the isolated clump to the L227 main cloud is, however, controversial.

\subsection{Properties of the L227 Cloud} 

The mean velocity of the L227 cloud, $\VLSR = 8.43$ \kms, 
gives the kinematic distance of either $D\!=\!1.7\pm 1.0$ kpc or $15.2\pm 1.0$ kpc 
if we use the Galactic rotation curve of Clemens (1985) with $R_0\!=\!8.5$ kpc. 
The errors came from possible random and streaming motions of $\pm 5$ \kms\ superposed on the Galactic rotation. 
At the far distance it is unlikely that associated HII region S32 would be visible in optical wavelength, 
and furthermore it would be 575 pc below the Galactic plane, 
over seven times the molecular cloud scale height of 40--75 pc (e.g., Sanders, Solomon, \&\ Scoville 1984). 
It must therefore be at the near distance. 
The kinematic distance is in good agreement with the distance to the stars exciting the HII region S32, 
$D\!=\!1.8\pm 0.6$ kpc, which was determined by spectrophotometry of the stars (Georgelin 1975). 
Hereafter we adopt $1.8$ kpc as the distance to the L227 cloud.

A virial theorem mass is determined by
\begin{equation}
M_{\rm VT} = 3 f_p \frac{S \sigma_V^2}{G} , 
\end{equation}
where $f_p \approx 2.9$ is a projection factor (Solomon et al. 1987), 
$\sigma_V$ is a velocity dispersion, and $S\!\equiv \! D {\rm tan}(\sqrt{\sigma_{\alpha}\sigma_{\delta}})$ 
is a size parameter. 
The size parameter and the velocity dispersion of the L227 cloud are 3.44 pc and 1.74 \kms\ respectively, 
giving a virial mass of $2.1\!\times\! 10^4 (f_p/2.9)$ $M_{\sun}$. The total CO luminosity defined by
\begin{equation}
L_{\rm CO} \equiv D^2 \int\!\!\!\int T_R^{*} d\Omega dV \;\; ({\rm K\:\kms\ pc^2}) 
\end{equation}
of the L227 cloud is $1.9\!\times\! 10^3$ K \kms\ pc$^2$. 
These cloud parameters closely follow the $S$-$\sigma_V$ and the $L_{\rm CO}$-$M_{\rm VT}$ relations of 
molecular clouds in the Galactic disk (Solomon et al. 1987). 
This means that the L227 cloud is a `normal' Galactic molecular cloud, 
and lends additional support to the adopted distance.

The molecular mass can also be estimated from the CO luminosity adopting the standard CO-to-H$_2$ conversion factor 
(e.g., Bloemen et al. 1986; Scoville et al. 1987; and Solomon et al. 1987), 
\begin{equation}
X\equiv N({\rm H}_2)/I_{\rm CO} = 3.0\!\times\!10^{20} \;\; {\rm (cm^{-2} (K\:\kms)^{-1})}, 
\end{equation}
where $N({\rm H}_2)$ is the H$_2$ column density and $I_{\rm CO}$ is the CO {\it J}=1--0 integrated intensity. 
Adopting the mean molecular mass of 2.72, we obtain $M\!=\!1.2\!\times\!10^4$ $M_{\sun}$, 
which agrees with the virial mass derived above within a factor of 2.

\section{DISCUSSION}

\subsection{Molecular Cloud/Pulsar Association ?} 

In Fig.4, we show X-ray images obtained with ASCA GIS (Kawai et al. 1998) 
with a contour map of integrated CO {\it J}=1--0 intensity. 
Two point-like sources are apparent in the X-ray image in the energy range from 0.5 to 10 keV (Fig.4{\it a}). 
The southern source corresponds to a young stellar cluster, 
while no optical counterpart of the northern source has been found. 
The northern source has a hard X-ray spectrum and is embedded in a nonthermally emitting X-ray nebula, 
exhibiting the characteristics of rotation-powered pulsars. 
We advance the following discussion assuming that the northern source 
originates from a rotation-powered pulsar, 
although pulsed emission from the source has not been detected yet.

The northeastern (NE) edge of the synchrotron nebula follows the rim of integrated CO emission 
and the SE extension of the synchrotron nebula nearly reaches the bottom of the molecular arc (Fig.4{\it c}). 
The NW edge of the synchrotron nebula corresponds to a dent in CO emission at $\VLSR\!=\!7$--12 \kms . 
This spatial anti-correlation between molecular gas and the synchrotron nebula can be attributed 
either to the absorption of soft X-ray photons by the L227 cloud or the interaction 
between the synchrotron nebula and the L227 cloud. 
Actually, with the standard CO-to-H$_2$ conversion factor, 
typical integrated intensity along the ridge of the L227 cloud, 
$I_{\rm CO}\!\geq\!30$ K \kms, corresponds to $N({\rm H})\!\geq\!2\!\times\!10^{22}$ cm$^{-2}$, 
which is sufficient to shield soft X-ray ($<\!2$ keV) photons. 
%Actually, the spatial anti-correlation seems to be clearer in the lower energy band image (Fig.4{\it b}) 
%though we do not have enough statistics. 

However, the possibility of the interaction between the synchrotron nebula and the L227 cloud can not be ruled out.
The spatial anti-correlation between the synchrotron nebula and molecular gas 
is also apparent even in the higher energy band image (Fig.4{\it c}), 
in which absorption by interstellar matter is not severe, 
which suggests that the synchrotron nebula could be interacting with molecular gas. 
The isolated clump at $\VLSR\!\simeq\! 19$ \kms\ (\S 3.1) 
could be evidence for dissociative J-shock (e.g., Hollenbach, \&\ McKee 1989) 
driven by the interaction with the synchrotron nebula,    
if it is really associated with the L227 main cloud.   
Here we examine the pressure balance between molecular gas and the synchrotron nebula. 
In general, the pressure inside of a molecular cloud is composed of both a turbulent and a thermal contribution, 
so that
\begin{equation}
p_{gas}/k = \rho \sigma_{turb}^2 + \frac{\rho}{\mu m_{\rm H}} T_k , 
\end{equation}
where $\rho$ ($=\!\mu m_{\rm H} n$) is the mass density, $k$ is the Boltzmann constant, 
$\sigma_{turb}$ is the turbulent velocity, $\mu$ is the mean molecular weight, 
$m_{\rm H}$ is the atomic mass unit, and $T_k$ is the kinetic temperature. 
The mean number density can be estimated by the column density, $N({\rm H}_2)\!=\!10^{22}$ cm$^{-2}$, 
which corresponds to $I_{\rm CO}\!=\!30$ K \kms, and assuming the depth of the cloud $\sim\!1$ pc, 
to be $\langle n({\rm H}_2)\rangle\!\simeq\!3\!\times\!10^2$ cm$^{-3}$. 
Using $\mu\!=\!2.72$ and $\sigma_{turb}\!=\!1$ \kms\ (from the CO linewidth), 
we find $p_{gas}/k\!\simeq\!10^5$ cm$^{-3}$ K. 
Thermal motion (the second term) is very small compared with the turbulent motion, 
contributing $\lesssim\! 10^4$ cm$^{-3}$ K to the pressure.

The relativistic pulsar wind flows outward until it terminates at a point $r_s$, 
where it comes into pressure equilibrium with the surrounding medium (Rees \&\ Gunn 1974). 
The pressure of the relativistic wind nebula at a radius $r_s$ is given by 
\begin{equation}
p_{wind} = \frac{\dot{E}}{4 \pi r_s^2 c}, 
\end{equation}
where we have assumed an isotropic steady wind and $\dot{E}$ is the energy input rate, 
which is equivalent to the spin-down luminosity ($L_{spin}$). 
The pulsar wind communicates its pressure to molecular gas mainly through the magnetic field. 
The pressure balance between the synchrotron nebula and molecular gas 
at $r_s\!=\!0.5 (\theta_s/1^\prime) (D/1.8\,{\rm kpc})$ pc requires a spin-down luminosity of
$L_{spin}\!\simeq\!10^{37} (\theta_s^\prime D_{1.8})^2 $ erg s$^{-1}$, 
which is similar to that of the Vela pulsar. 
This $L_{spin}$ and the X-ray luminosity of the source including the nebula, 
$L_X\!\simeq\!10^{33}$ erg s$^{-1}$,
roughly follows the $L_{spin}$-$L_X$ trend found for the Galactic rotation-powered pulsars (Saito 1997). 
If the empirical $L_{spin}$-$L_X$ relation strictly holds, 
$L_X\!\simeq\!10^{33}$ erg s$^{-1}$ leads to $L_{spin}\!\simeq\!10^{36}$ erg s$^{-1}$, 
suggesting a pulsar wind shock at $\theta_s\!\sim\!20^{\prime\prime}$ for our adopted nebula pressure. 
We concluded that the synchrotron nebula is roughly in pressure equilibrium with, 
or has suppressed its expansion by molecular gas at the contact surface. 
Planned {\it AXAF} observations may further test this, by measuring the scale of the nebular wind shock.
In the latter case the arc-shape of the L227 cloud may have been formed 
by ionization of molecular gas by the young stellar cluster in S32.

If the pulsar is really associated with the L227 cloud, it is 70 pc below the Galactic plane. 
Depending on the pulsar age an origin in the massive star association to the south is plausible. 
However, the lack of an obvious type II supernova remnant 
in the vicinity also makes a more distant origin possible. 
With a typical 1-D birth velocity of $\sigma_V\!\simeq\!300$ \kms\ (Lorimer et al. 1997) 
the transit time from the young star cluster associated with the soft X-rays to the south is $\sim\!10^4$y, 
while travel from the Galactic plane would imply a pulsar age of $\sim\!2\times 10^5$y. 
Recently a large-scale CO survey with the NANTEN telescope revealed that 
molecular gas in this region seems to form a large expanding-shell-like structure with a radius of $\sim\!5$--$10$ pc, 
and that the L227 cloud corresponds to a Galactic northeastern portion of it (Yamaguchi et al. 1998).  
The formation of the large expanding-shell-like structure by a type II supernova or a series of supernova explosions  
implies a pulsar age of $\sim\!(2$--$3)\times 10^4$y.

\subsection{$\gamma$-ray Emission from the Pulsar/Molecular Cloud System} 

The X-ray spectrum of the synchrotron nebula   
requires a power law energy distribution for electrons to $\sim$ TeV energies.   
These electrons may have been injected as pulsar winds 
and be accelerated further by the first-order Fermi mechanism in wind termination shock   
(e.g., Kennel \&\ Coroniti 1984a; 1984b).    
These high energy electrons would contribute to high energy $\gamma$-ray emission from the system.  
In the case of the Crab nebula, it is suggested that synchrotron spectrum extends up to 100 MeV 
while inverse Compton (IC) spectrum excels in GeV energies (De Jager et al. 1996; Atoyan \&\ Aharonian 1996).   
%and that the lower energy ($70\,\mbox{MeV}\!\leq\!h \nu \!\leq\!150\,\mbox{MeV}$) $\gamma$-ray flux shows nominal time variablity .  
In contrast, the $\gamma$-ray spectrum of 2EG J1811--2339 can be fitted 
by a single power law (Merk et al. 1996) 
which can not match to the extension of the synchrotron spectrum.  
This suggests that the synchrotron spectrum has a cut off between 
$10\,\mbox{keV}\!\leq\!h \nu \!\leq\!50\,\mbox{MeV}$
and that another radiation mechanism produces GeV photons.

If high energy electrons in pulsar winds can move into the L227 cloud, 
dense interstellar matter in the L227 cloud can be responsible for 
the $\gamma$-ray emission from 2EG J1811--2339 through relativistic bremsstrahlung.   
The $\gamma$-ray flux from such a system could be time variable 
because of instabilities in the interaction layer and of inhomogeneous distribution of target baryons.  
The variation time scale would be a few years 
since the electron flux would propagate with nearly the light speed 
across the dense interstellar matter with the inhomogeneity on a scale of $\lesssim\!0.5$ pc.
Here we discuss the $\gamma$-ray emission model for 2EG J1811--2339 based on the above scenario 
and then compare the model spectrum with the observed spectrum. 
We also examine IC process on the same condition as an alternative possibility, 
although it is expected to be steady.
Details of the model calculations will be presented in the other paper (Naito et al. 1999).

Primary momentum spectrum of electrons accelerated in shocks 
by first-order Fermi mechanism is assumed as 
\begin{equation}
\frac{dN}{dP} = N_{0} \left( \frac{P}{m c}\right)^{-\alpha} 
{\rm exp}\left(-\frac{P}{P_{\rm max}}\right) \ \ \ \ (P > P_{\rm inj})
\ \ \ \ ({\rm cm}^{-3} \ {\rm (eV/}c{\rm )^{-1}}) 
\end{equation}
(Drury 1983; Blanford \& Eichler 1987) 
where $m$ is the electron mass and $c$ is the light speed.  
We presume that electrons with $P\!>\!P_{\rm inj}$ are injected into 
the acceleration process to have the spectrum shown in equation (6) 
and that those with $P\!<\!P_{\rm inj}$ have constant momentum distribution.  
In this paper, we allow that the momentum spectrum of electrons is invariable 
when they propagate from acceleration site to the X-ray nebula and the L227 cloud.

Assuming that the X-ray flux from the system observed with ASCA 
is purely due to synchrotron radiation
induced by electrons having spectrum expressed by equation (6), 
$\alpha\!=\!2.2$ is derived by the photon index ($\beta\!=\!1.6$) and 
$N_{\rm e}/c\!=\!2.4 \times 10^{- 10}\ (D/1.8 {\rm kpc})^{2} (r_s/1 {\rm pc})^{-3}\ 
(B/10 \mu{\rm G})^{- (\alpha+1)/2}\
{\rm {cm}^{-3} \ {eV}^{-1}}$ by the intensity, 
where $B$ and $r_s$ are the magnetic field strength and the length scale of the X-ray emitting region, respectively.  
The synchrotron spectrum is calculated on the basis 
of $\delta$-function approximation (Rybick \& Lightman 1979) 
that electron with energy $E_{\rm e}$ emits characteristic 
photon with energy $h\nu\!=\!15\,(E_{\rm e}/{\rm TeV})^2\,(B/{\rm G})$ keV.  
The observed ASCA and EGRET spectra require that the flux breaks at 
$10\,\mbox{keV}\!\leq\!h\nu\!\leq\!50\,\mbox{MeV}$.  
The hypothesis that $B^2/(8 \pi)\!\sim\!p_{gas}$ (see \S 4.1) leads to $B\!\sim\!2 \times 10^{- 5}$ G,  
and then $1.8 \times \!10^{14}\,\mbox{eV}\!\leq\!E(P_{\rm max})\!\leq\!1.3 \times \!10^{16}\,\mbox{eV}$.  
We fix conservatively as $E(P_{\rm max})\!=\!1 \times \!10^{15}$ eV.  
In our model, $E(P_{\rm inj})$ should be equivalent with a typical energy of pulsar wind particle,  
of which we don't have enough information.   
We thus chose $E_{\rm kin}(P_{\rm inj})$ 
so that the calculated spectrum is consistent with the observed X-ray and $\gamma$-ray fluxes.

The X-ray to $\gamma$-ray spectrum is calculated based on 
the synchrotron, bremsstrahlung, and inverse Compton (IC) models.  
We assume that synchrotron and IC photons arise from the X-ray nebula only 
while the bremsstrahlung $\gamma$-rays come from entire region of the L227 cloud.  
In the bremsstrahlung model, 
we use a cross section including screening effects (Koch \& Motz 1959), 
which is applicable to molecular cloud with low ionization.  
In the IC model, 
a cross section including relativistic effects (Jones 1968) is employed. 
We consider the 2.7 K CMB and galactic background radiation in infrared wavelength as target photons 
using the diluted black body radiation model (Mathis, Merzger, \& Panagia 1983).
We neglect the contributions of specific photons 
since they are difficult to estimate (Gaisser, Protheroe \& Stanev 1998).  
When the synchrotron flux 
$F_{sync}$ is fixed, 
the fluxes due to the other processes are scaled by parameters as, 
$F_{brem}\!\propto\!B^{- (\alpha+1)/2}\,\overline{n}\ r$,  
$F_{ic}\!\propto\!B^{- (\alpha+1)/2}$, 
where $\overline{n}$ is the mean target proton density, 
and $r$ is the morphologic conversion factor from synchrotron emitting region to bremsstrahlung emitting region. 
According to the conservation of particle flux, 
the bremsstrahlung flux is typically proportional to 
$(V_{brem}/ V_{sync})\,(S_{sync}/ S_{brem})\!\sim\!r$, 
where $V_{brem}$,  $V_{sync}$, $S_{brem}$, and $S_{sync}$ 
indicate typical volumes and cross sections of each emitting area, respectively.

Fig.5 shows the best-fit spectrum to the observed X-ray and $\gamma$-ray spectra.  
We assume $B\!=\!20\,\mu$G, $r\!=\!10$, $l\!=\!3$ pc, 
$E_{e, kin}\!=\!\sqrt{(P_{inj} c)^2+({m_{\rm e} c^2})^2}-m_{\rm e} c^2\!=\!10^7$ eV 
and $p_{gas}\!=\!1.0\times 10^{-11}$ erg cm$^{-3}$. 
The best-fit value for the mean target proton density 
is $\overline{n}\!=\!2.5\times 10^3 $ cm$^{-3}$, 
which agrees with the value obtained in the preceding section in a factor of 3, 
$\overline{n}\!\simeq\!\mu \langle n({\rm H}_2)\rangle\!=\!8\times 10^2$ cm$^{-3}$ (\S 4.1). 
This discrepancy could be due to blind application of the standard CO-to-H$_2$ conversion factor, 
or smaller magnetic field strength in the synchrotron nebula than the assumed. 
In this calculation, the electron pressure of 
$P_{\rm e} \simeq 3.0 \times 10^{-11} (D/1.8 {\rm kpc})^2
(r_{\rm s }/1 {\rm pc})^{-3} (B/20 {\rm \mu G})^{-1.6}
(E(P_{\rm inj})/10^7 {\rm eV})^{-0.2}\ \ {\rm erg\ cm^{-3}}$
is required,  which is almost consistent with the gas
pressure of the molecular cloud given by equation (4)  if
$\overline{n}\!=\!2.5\times 10^3 $ cm$^{-3}$. Relativistic
bremsstrahlung dominates the $\gamma$-ray emission  in the
EGRET band from this system,  while IC scattering process
makes a negligible contribution.  Although there are
uncertainties in particle propagation and penetration
processes,  our scenario reproduces well the X-ray to
$\gamma$-ray spectrum from 2EG J1811--2339.

From fig.5 one can see that the TeV $\gamma$-ray flux exceeds the sensitivity of 
ground based \v{C}erenkov telescope  
[$E F(E)\!\gtrsim\!1\,\mbox{cm}^{-2}\mbox{s}^{-1}\mbox{eV}$] by one order.  
The detection of TeV $\gamma $-rays could confirm the existence of system 
that pulsar wind interacts with molecular cloud in 2EG J1811--2339. 
In this case we expect a spectrum, of which index $\sim\!2.2$, 
extending up to $h\nu\!\geq\!180/\sqrt{(B/20 \mu {\rm G})}$ TeV 
with emission spreading over the L227 cloud, $\theta\!\sim\!10'$.  
An IC origin would have emission concentrated at the synchrotron nebula core; 
pulsar magnetospheric emission should not produce strong TeV flux.

\section{CONCLUSIONS}

We have presented CO {\it J}=1--0 mapping data of the dark nebula Lynds 227, 
which is possibly associated with the unidentified EGRET source 2EG J1811--2339. 
The major results are the following:

\begin{enumerate}

\item Molecular gas closely follows the dark nebula Lynds 227 seen in obscuration of the background star light. 
The L227 cloud shows spatial anti-correlation with the X-ray synchrotron nebula, 
suggesting that the synchrotron nebula is interacting with the Lynds 227 cloud. 
The mass of the L227 cloud is about (1--2)$\times 10^4$ $M_{\sun}$. 

\item The X-ray synchrotron nebula is roughly in pressure equilibrium with, 
or has suppressed its expansion by molecular gas in the L227 cloud. 
If the X-ray synchrotron nebula is in equilibrium with molecular gas, 
an embedded pulsar with a spin-down luminosity of $\sim 10^{37}$ erg s$^{-1}$ is required.

\item We propose a $\gamma$-ray emission mechanism for 2EG J1811--2339 
in which high energy charged particles accelerated in shocks  
collide with dense interstellar matter, 
and thereby generate high energy $\gamma$-ray photons mainly through relativistic bremsstrahlung. 
This scenario can explain the observed $\gamma$-ray flux as well as its time variability. 
Planned high resolution X-ray and TeV observations can confirm this hypothesis. 

\end{enumerate}

\acknowledgments
We are grateful to the NRO staff for excellent support in the observations. 
We also thank Dr. Bradley C. Rubin for checking the English. 
T.O. is financially supported by the Special Postdoctoral Researchers Program of RIKEN. 
T.N. is supported by the National Astronomical Observatory as COE (Center of Excellence) research fellow. 
R.W.R. is supported in part by NASA grant NAG 5--3263.

\clearpage

\figcaption[f1.eps]{
Contour map of CO {\it J}=1--0 intensity integrated over the velocity range $\VLSR\!=\!0$--$25$ \kms\ 
of the area covered by the coarse grid survey. 
The contours begin at 10 K \kms\ and thereafter are spaced by 5 K \kms\ intervals. 
White contours begin at 40 K \kms.
The 95 \%\ uncertainty circle for the 2EG J1811--2339 position, 
determined based on photons with energies greater than 1 GeV (Lamb \&\ Macomb 1997), is also shown.
Hatched area and crosses denote the locations of the X-ray synchrotron nebula and point sources, respectively. 
Broken rectangle shows the area covered by the fine grid survey. 
\label{fig1}}

\figcaption[f2.eps]{
Contour map of CO {\it J}=1--0 intensity integrated over the velocity range 
$\VLSR\!=\!0$--$25$ \kms\ of the area covered by the fine grid survey. 
The contours are drawn at every 5 K \kms. 
White contours begin at 40 K \kms. 
%({\it b}) Map of $^{13}$CO {\it J}=1--0 intensity integrated over the velocity range %$\VLSR\!=\!0$--$25$ \kms.
%The contours are drawn at every 2 K \kms. 
%The region covered by $^{13}$CO map is shown in the CO map by a dotted rectangle. 
\label{fig2}}

\figcaption[f3.eps]{
Velocity channel maps of CO {\it J}=1--0 emission integrated over successive 2.5 \kms\ widths. 
Contours are set at intervals of 2.5 K \kms. White contours begin at 22.5 K \kms. 
\label{fig3}}

\figcaption[f4.eps]{
The smoothed X-ray images obtained with ASCA GIS, covering energy ranges ({\it a}) 0.5--10 keV, 
({\it b}) 0.5--2.2 keV, and ({\it c}) 2.2--10 keV, respectively. 
Contour map of CO {\it J}=1--0 intensity integrated over the velocity range $\VLSR\!=\!0$--$25$ \kms\ is superposed. 
The gray depth is scaled between minimum and maximum intensity of each image. 
The CO contours are drawn at every 5 K \kms. 
\label{fig4}}

\figcaption[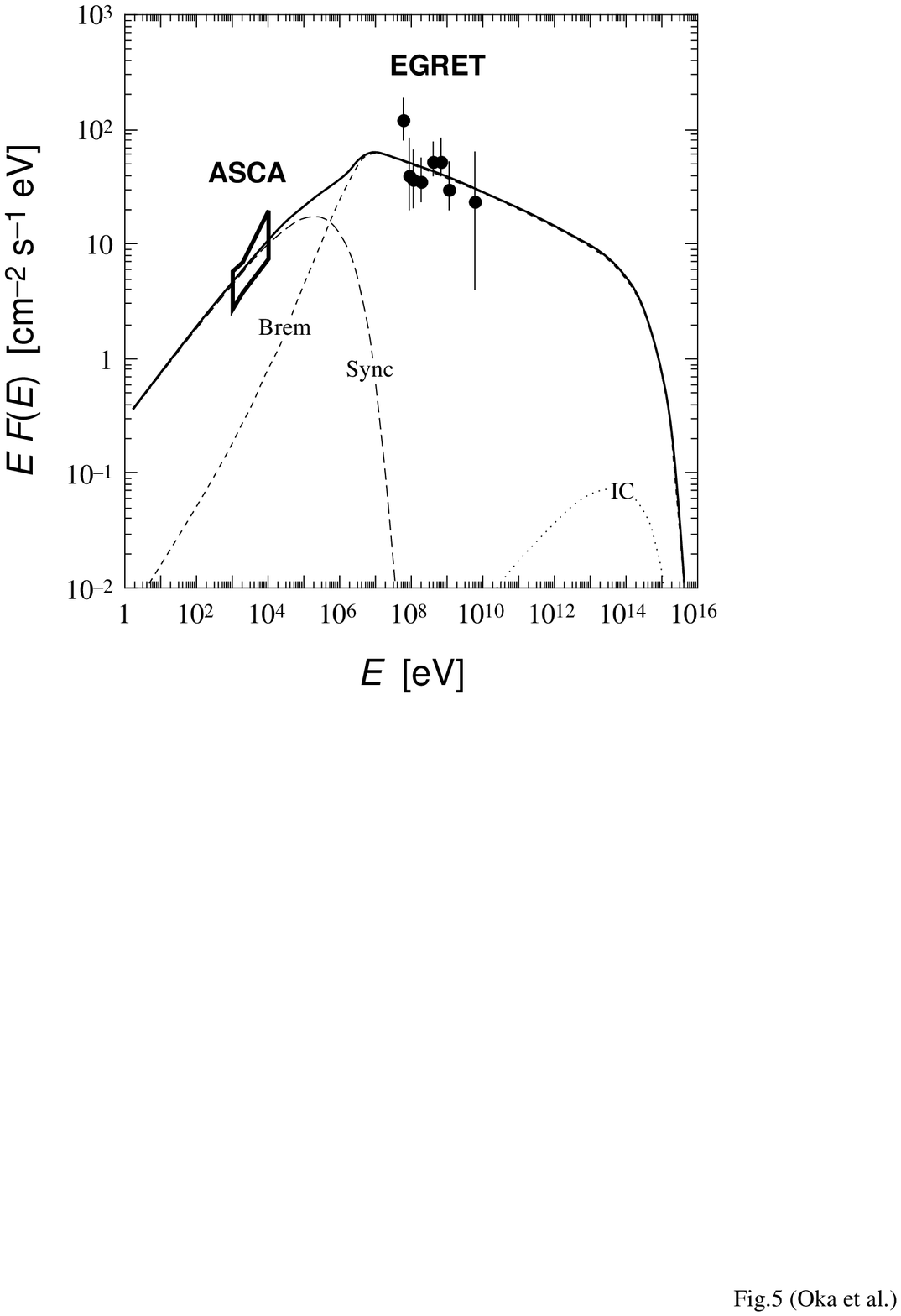]{
Observed X-ray and $\gamma$-ray spectra of 2EG J1811--2339. 
X-ray data are from Kawai et al. (1998) and $\gamma$-ray data from Merck et al. (1996). 
Fits to the data are superposed: long-dashed line, synchrotron (Sync); 
mid-dashed line, relativistic bremsstrahlung (Brem); 
short-dashed line, inverse Compton scattering (IC); 
thick solid line, sum of all processes. 
\label{fig5}}

\end{document}